\documentclass[preprint,onecolumn]{revtex4}
\usepackage{graphicx}

\begin{document}

\title{Cosmic ray-driven bioenergetics for Life in Molecular Clouds}
\author{Lei~Feng}
\email{fenglei@pmo.ac.cn}
\affiliation{Key Laboratory of Dark Matter and Space Astronomy, Purple Mountain Observatory, Chinese Academy of Sciences, Nanjing 210023}
\affiliation{School of Astronomy and Space Science, University of Science and Technology of China, Hefei, Anhui 230026, China}
\affiliation{Joint Center for Particle, Nuclear Physics and Cosmology,  Nanjing University -- Purple Mountain Observatory,  Nanjing  210093, China}

\begin{abstract}
According to models such as panspermia or the Nebula-Relay hypothesis, the ancestors of life on Earth once lived in molecular clouds. Then what are the energy source and bioenergetics for such lifeforms? In this paper, we propose a new bioenergetic mechanism powered by cosmic ray ionization of hydrogen molecules and we argue that it may relate to or be the origin of chemiosmosis. Based on this mechanism, we suggest that the Last Universal Common Ancestor is a type of lifeform that utilize hydrogen molecules as donors of electron transport chains.\end{abstract}
\maketitle
\section{Introduction}

Life's genetics and biological activity rely heavily on its bioenergetics which governs the energy flow of living systems. These include transforming energy, producing and utilizing adenosine triphosphate (ATP) molecules, etc. These steps effectively convert chemical energy into ATP, essential for life. On Earth, cellular respiration occurs in two ways: aerobic and anaerobic respiration.
Peter D. Mitchell proposed the chemiosmotic theory in 1961 \cite{Mitchell} and pointed out that ATP synthesis is driven by an electrochemical gradient established across the plasma membrane as hydrogen ions (protons) diffuse from areas of high proton concentration to those of lower concentration. The energy released through the electron transport chain maintains the proton concentration gradient as a proton pump.

The ancestors of Earth's life might be lived in molecular clouds according to some models such as panspermia and Nebula-Relay hypothesis \cite{nebula-relay} which suggested that the ancestors of Earth's life may have originated in Sun's predecessor planetary system and lived in the pre-solar nebula (a molecular cloud where the solar system was formed). According to the Nebula-Relay hypothesis, life in the solar system origins on the planet of the Sun's predecessor star through complex physicochemical interactions which is similar to the abiogenesis on Earth \cite{Oparin,Haldane}.
Then, these primitive organisms populated in the pre-solar nebula created after the death of this predecessor star. Compared to the panspermia theory \cite{Panspermia}, this model suggests that homologous lifeforms or their fossils can be found throughout the solar system. In Ref. \cite{Chirality}, the author found that the ultra-low temperature environment of molecular clouds might explain the origin of the chirality of biological molecules.

How would life get enough energy in molecular clouds? What is the bioenergetics for life in molecular clouds? As proposed in Ref. \cite{nebula-relay}, cosmic ray (CR) is the primary energy input of molecular clouds and may also be the source of life there.
CR particles possess a great deal of energy, and their collisions with molecules in molecular clouds induce the production of lower-energy particles, thus reducing the potential damage caused by CR. This paper proposes a new bioenergetics mechanism, which is driven by CR ionization of hydrogen in molecular clouds. It is worth noting that this scenario applies to all lifeforms in molecular clouds, not just the panspermia and Nebula-Relay hypothesis.

This paper is organized as follows: In Sec.2, the author introduces a bioenergetics mechanism driven by the CR ionization of hydrogen and argues that may also be the origination of chemiosmosis. Possible testing experiments and discussions are summarized in the final section.

\section{Cosmic ray and the ionization of hydrogen molecules}

CRs are high-energy particles that travel through the galaxy or universe at nearly the speed of light discovered by Victor Hess in 1912. The origin of CR in the Milky Way is usually considered to be the supernova explosion. Otherwise, the source of ultra-high-energy CRs remains a mystery. The energy spectrum of CRs is roughly a power law spectrum with different spectral indices in different energy ranges. Precise measurements of the CR energy spectrum can help us understand the properties of the cosmic-ray sources, their transition in the universe and so on. There are already lots of CR measurement experiments, such as AMS-02 \cite{ams-02}, Fermi-LAT \cite{feimi-lat}, PAMELA \cite{pamela}, DAMPE \cite{dampe}, Calet \cite{calet} and so on. 


There are many forms of interaction between cosmic rays and the interstellar medium. But the CR ionization of hydrogen is the main interaction for low-energy cosmic rays in molecular clouds and it mainly involves the following processes 
\begin{eqnarray}
{\rm CR} + {\rm H_2} &\rightarrow& {\rm CR} + {\rm H^+_2} + {\rm e}, \\
{\rm CR} + {\rm H_2} &\rightarrow& {\rm CR} + {\rm H} + {\rm H^+} + {\rm e},\\
{\rm CR} + {\rm H_2} &\rightarrow& {\rm CR} + {\rm 2H^+} + {\rm 2e},
\end{eqnarray}
where ${\rm CR}$ denotes the charged CR particles and secondary particles produced by the collision between CR and molecular cloud, such as protons, heavy nuclei, electrons, positrons, etc. If the produced electrons have suﬃciently energy, it would induce further ionizations of hydrogen molecules as its relatively short range.

It should be pointed out that the formation of $\rm H_2^+$ (i.e. the process in Eq.(1)) is the major process and the production rate of electrons per $\rm H_2$ can be described as \cite{Padovani:2009bj}
\begin{equation}
\zeta^e=4\pi\sum_k\int_{I({\rm H_2})}^{E_{\rm max}}j_k(E_k)[1+\phi_k(E_k)]\sigma_k^{\rm ion}(E_k){\rm d}E_k,
\end{equation}
where $j_k(E_k)$ is the number of CR particles per unit area, time, solid angle and energy for CR component $k$, $I({\rm H_2})$ is the ionization potential of the hydrogen molecule, $E_{\rm max}$ is the maximum energy considered and {\bf $\sigma_k^{\rm ion}$ is the cross section for ionization of molecular hydrogen}. The quantity $\phi_k(E_k)$ in the above equation is a correction factor accounting for the ionization of hydrogen molecules by secondary electrons, which is described as \cite{Padovani:2009bj}
\begin{equation}
\phi_k(E_k) \equiv \frac{1}{\sigma_k^{\rm ion}(E_k)}\int_{I({\rm H_2})}^{E_{\rm max}^\prime}P(E_k,E_{\rm e}^\prime)\sigma_{\rm e}^{\rm ion}(E_{\rm e}^\prime){\rm d}E_{\rm e}^\prime,
\end{equation}
where $P(E_k, E_{\rm e}^\prime)$ is the probability that electrons with energy $E_{\rm e}^\prime$ are produced from the ionization of CR particles with energy $E_k$. 
More detailed calculations can be found in Ref. \cite{Padovani:2009bj}.

High-energy photons, such as X-rays and gamma rays, can also knock electrons out of hydrogen molecules through the photoelectric effect. Here we do not discuss such processes in this draft.

\section{Cosmic ray-driven bioenergetics and the origin of chemiosmosis}

Molecular cloud is a type of interstellar cloud. And its average density is about $\rm 10^2 \sim 10^4$ molecules per cubic centimetre, which is most commonly molecular hydrogen. The typical temperature of a molecular cloud is about $\rm 10\sim 20$ Kelvin (Kelvin is a unit of temperature and Celsius = Kelvin - 273.15).  Although hydrogen's melting (boiling) point is 13.99~K (20.27~K), hydrogen maintains in the gaseous phase because of its low pressure. We suppose that the molecular cloud life has a cell-like membrane structure. If the hydrogen can be enriched in a cell, then the hydrogen pressure is also enlarged and hydrogen may maintain a liquid state in molecular cloud life. If so, the internal environment of molecular cloud life is liquid hydrogen, similar to the liquid water environment of cells on Earth. That will be very favourable for the biochemical reactions therein.

CR is a vital type of energy injection and responds to both the ionization and heating of molecular clouds. It is also a critical component in the chemistry and dynamics of the interstellar medium (see Ref.~\cite{Dalgarno2006} for more information). The nature of bioenergetics is to gradually release the energy carried by electrons through the electron transport chain to maintain life activities. Moreover, the CR ionization process could produce many electrons with suitable energy. So we propose that the CR-induced ionization of hydrogen power the electron transport chain of lifeforms in molecular clouds. Electrons may not enter the electron transport chain directly but first interact with other molecules, such as nicotinamide adenine dinucleotide ($\rm NAD^+$). It is more natural for electrons produced by CR ionization to enter the electron transport chain directly based on the principles of simplicity and optimal energy utilization. However, the electron donor is still molecular hydrogen and the energy essentially comes from the ionization process of CR with or without an intermediate electron transporter. Furthermore, the authors in Ref. \cite{Padovani:2009bj} found that the ﬂux of either CR electrons or protons increases at low energies that would produce much more suitable electrons for the electron transport chain.

The produced ion ${\rm H^+_2}$  from CR ionization of hydrogen molecules can be depleted rapidly by reacting with molecular hydrogen as follows\cite{ionreaction}
\begin{equation}
{\rm H^+_2} + {\rm H_2} \rightarrow {\rm H^+_3} + {\rm H}.
\end{equation}
Then ${\rm H^+_3}$ is removed by reacting with the neutral molecular through proton transfer reactions
\begin{equation}
{\rm H^+_3} + {\rm X} \rightarrow {\rm HX^+} + {\rm H_2},
\end{equation}
where ${\rm X}$ is the neutral molecular in nebula, such as ${\rm CO}$, ${\rm H_2O}$, ${\rm N_2}$ and so on. The products of the reaction between ${\rm H^+_2}$ ions and carbon compounds may be crucial for synthesizing the organic molecules of lifeforms in molecular clouds. The hydrogen atoms are further ionized to $\rm H^+$ or form hydrogen molecules through a grain-catalyzed reaction
\begin{equation}
{\rm H} + {\rm H} \rightarrow {\rm H_2}.
\end{equation}
In general, the cellular fluid of molecular cloud life is a mixture of hydrogen molecules, hydrogen atoms and protons.

Though parts of energetic secondary electrons may not be directly involved in the electron transport chain, the further ionization of ${\rm H_2}$ and hydrogen atom by secondary electrons reduces the energy of electrons and enlarges the number of electrons and protons simultaneously. Such a process maximizes the utilization efficiency of high-energy CRs. Such further ionization processes end until the electron's energy is suitable for entering the electron transport chain and pumping protons into the intermembrane space.

\begin{figure}
\centering
\includegraphics[width=150mm,angle=0]{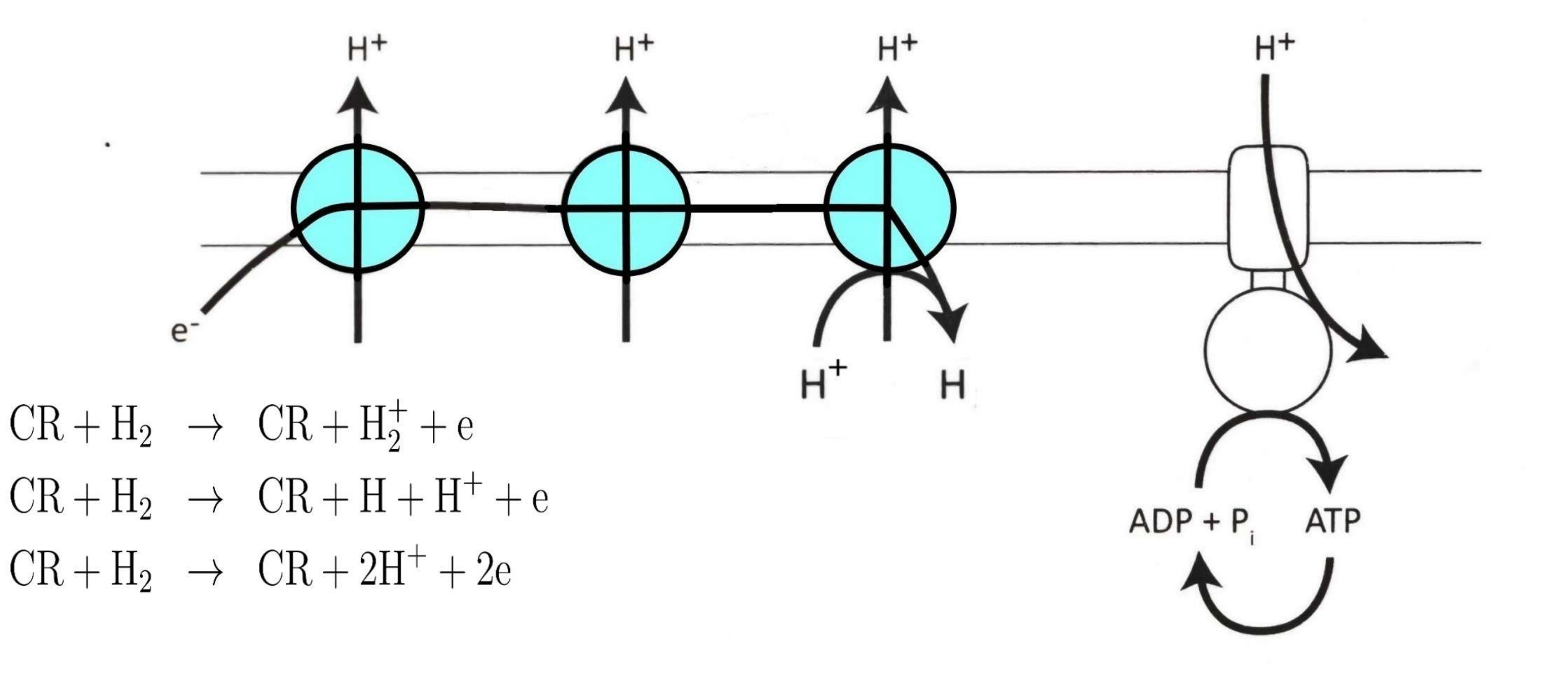}
  \caption{Schematic diagram of the electron transport chain powered by the CR ionization of hydrogen where ATP and ADP are the abbreviation of adenosine triphosphate and adenosine di-phosphate respectively.}
  \label{fig:chain}
\end{figure}

Like lifeforms on Earth, the electrochemical proton gradient drives ATP synthesis. The flow of electrons then combines with ${\rm H^+}$ to form hydrogen atoms acting as donors and acceptors. Of course, there may be other acceptors, and we will not discuss them in detail here. A schematic representation of this process is shown in Fig. \ref{fig:chain}. Hydrogen molecules are the most abundant molecules in molecular clouds, and their ionization and subsequent chemical reaction naturally induce the production of atoms/protons. Given the scarcity of organic compounds and the abundance of hydrogen molecules in molecular clouds, life probably developed a proton gradient to drive the production of energy usable for their survival. This mechanism may relate to or even be the origin of chemiosmosis.

As part of their energy is transformed and utilized by lifeforms in molecular clouds, the energy spectrum of CR is expected to be slightly modified. The exact extent of this impact depends on the density of lifeforms and energy utilization efficiency. We plan to investigate this topic further in the future. In addition, many other factors (e.g., propagation parameters, the flux of primary cosmic rays, and the density and distribution of molecular clouds) can also influence the energy spectrum of cosmic rays. It would be technically challenging to distinguish the influence of lifeforms from so many complicated factors.

\section{Some discussions}

The CR-driven bioenergetics discussed in this paper is essentially a reversible redox reaction as follows
\begin{equation}
{\rm H_2} \rightleftharpoons {\rm 2H^+}+2e^-.
\end{equation}
This process is similar to the bacteria that use hydrogen as an energy source catalyzed by hydrogenases. The ability to metabolize hydrogen is widespread and appears inherent to Earth's life. More than 30\% of known microbial taxa possess hydrogenase genes \cite{peters}.  Furthermore, it has been reported that about 70\% of gastrointestinal microbial species listed in the Human Microbiome Project contain the genetic information necessary for metabolizing hydrogen molecules \cite{Wolf}.
If the Earth's atmosphere was initially rich in hydrogen, it is assumed that hydrogenases naturally evolved to utilize molecular hydrogen as an energy source. Hydrogen molecules have been identified as electron donors in hydrothermal vent animal symbioses \cite{Petersen}.

Suppose that the last universal common ancestor (LUCA) came from the pre-solar nebula as suggested by the Nebula-Relay hypothesis and was driven by the cosmic ray ionization of hydrogen molecules. Then LUCA is related to hydrogen bacteria and utilize molecular hydrogen as electron donor. In Ref. \cite{Madeline}, the authors found that LUCA may be a heat-loving microbe feeding on hydrogen gas by surveying nearly two thousand genomes of modern microbes. 
In Ref \cite{Lane}, the authors argued that the first donor of LUCA was hydrogen which is also consistent with our model. However, the electron acceptor is carbon dioxide in Ref. \cite{Madeline,Lane} which is different from the mechanism we discussed here. 

Testing this bioenergetic mechanism is challenging. However, one possible way is simulating the bioenergetic process of molecular cloud life by putting microorganisms closest to LUCA into liquid hydrogen, irradiating them with high-energy particles and studying their biochemical process.



\section{Summary}

In this draft, the author discusses the bioenergetics of life in molecular clouds and introduces a new bioenergetic mechanism driven by CR ionization of hydrogen. In this model, harmful CR radiations are transformed into bio-available energy.
The main prediction of this model is that LUCA is a type of microbe which use hydrogen molecules as donors of electron transport chains. Protons, generated from CR ionization and subsequent chemical reactions, can be used for transmembrane transport and ATP formation, potentially explaining the origin of chemiosmosis. Life in molecular clouds could gain sufficient energy to sustain its activities.

\acknowledgments
We thank Yu-Xing Cui for his generous help in perfecting this article.
This work is supported by the National Key R\&D Program of China (Grants No. 2022YFF0503304), the National Natural Science Foundation of China (12220101003), National Natural Science Foundation of China (Grants No. 11773075) and the Youth Innovation Promotion Association of Chinese
Academy of Sciences (Grant No. 2016288).




\renewcommand{\refname}{References}

\end{document}